\documentclass{sig-alternate}
\usepackage{cite}
\usepackage{graphicx}

\usepackage[utf8]{inputenc}
\usepackage{amsmath}
\usepackage{amsfonts}
\usepackage{amssymb}
\usepackage{algpseudocode}
\usepackage{algorithm}
\usepackage{subcaption}

\usepackage{courier}
\usepackage{listings}

\begin{document}
\conferenceinfo{Middleware'15, } { December 07-11, 2015, Vancouver, BC, Canada} 
 \CopyrightYear{2015}
 \crdata{ISBN 978-1-4503-3618-5/15/12} 
 \clubpenalty=10000 
 \widowpenalty = 10000

\title{{\ttlit R-Storm}: Resource-Aware Scheduling in Storm}

%
%
%
%
%

\numberofauthors{5} 
%
\author{
%
%
\alignauthor Boyang Peng\\
       \affaddr{University of Illinois, Urbana-Champaign}\\
       \email{bpeng@illinois.edu}
\alignauthor Mohammad Hosseini\\
       \affaddr{University of Illinois, Urbana-Champaign}\\
       \email{shossen2@illinois.edu}
\alignauthor Zhihao Hong\\
       \affaddr{University of Illinois, Urbana-Champaign}\\
       \email{hong64@illinois.edu}
\and  
\alignauthor Reza Farivar\\
       \affaddr{Yahoo! Inc.}\\
       \email{rfarivar@yahoo-inc.com}
\alignauthor Roy Campbell\\
       \affaddr{University of Illinois, Urbana-Champaign}\\
       \email{rhc@illinois.edu}
}


\maketitle
\begin{abstract}
The era of big data has led to the emergence of new systems for real-time distributed stream processing, e.g., Apache Storm is one of the most popular stream processing systems in industry today. However, Storm, like many other stream processing systems lacks an intelligent scheduling mechanism. The default round-robin scheduling currently deployed in Storm disregards resource demands and availability, and can therefore be inefficient at times.   We present R-Storm (Resource-Aware Storm), a system that implements resource-aware scheduling within Storm.  R-Storm is designed to increase overall throughput by maximizing resource utilization while minimizing network latency. When scheduling tasks, R-Storm can satisfy both soft and hard resource constraints as well as minimizing network distance between components that communicate with each other. We evaluate R-Storm on set of micro-benchmark Storm applications as well as Storm applications used in production at Yahoo! Inc.  From our experimental results we conclude that R-Storm achieves 30-47\% higher throughput and 69-350\% better CPU utilization than default Storm for the micro-benchmarks. For the Yahoo! Storm applications, R-Storm outperforms default Storm by around 50\% based on overall throughput. We also demonstrate that R-Storm performs much better when scheduling multiple Storm applications than default Storm.

\end{abstract}

\category{1.1}{Middleware for emerging cloud computing platforms}{}
\category{1.2}{Middleware for data-intensive computing (Big Data)}{}
\category{2.5}{Real-time solutions and quality of service}{}

\terms{Distributed Computation, Scheduling}

\section{Introduction}
As our society enters an age dominated by data, processing large amounts of data in a timely fashion has become a major challenge. As of 2012, 2.5 \textit{exabytes} $(2.5\times 10^{18})$ of data were created every day \cite{bigdata2012}, which increased to 2.3 zettabytes $(2.3\times 10^{21})$ of data as of 2014 \cite{bigdata2014a}. This number is projected to grow rapidly in the next few years as smart devices such as smartphones become more and more popular.

In the past decade, distributed computation systems such as \cite{hadoop_web}\cite{hive_web}\cite{pig_web}\cite{spark_web}\cite{isard2007dryad} have been widely used and deployed to handle big data. Many systems, like Hadoop, have a batch processing model which is intended to process static data.  However, a demand has arisen for frameworks that allow for processing of live streams of data and answering queries quickly.  Users want a framework that can process large dynamic streams of data on the fly and serve results to potential customers with low latency.

Storm is a distributed computation system built to address this concern. Storm is an open source distributed real-time computation system, for which the goal is to reliably process unbounded streams of data in an easy to program framework. What Hadoop has done for Big Data batch processing \cite{storm_web}, Storm is effectively poised to do for real-time processing.

Currently, the Storm platform uses pseudo-random round robin task scheduling and task placement on physical machines. This default scheduling algorithm is simplistic and not optimal in terms of throughput performance and resource utilization.  Default Storm does not consider resource availability in the underlying cluster or resource requirement in the of Storm topologies when scheduling. A Storm application or topology (defined in Section~\ref{overview}) is a user-defined application that can have any number of resource constraints for it to run.  

Not considering resource demand and resource availability when scheduling can be problematic.  The resource on machines in the cluster can be easily over-utilized or under-utilized which can cause problems ranging from catastrophic failure to execution inefficiency.   For example, a Storm cluster can suffer a potentially unrecoverable failure if certain executors attempt to use more memory than is available.  Over-utilization of resource other than memory can also cause the execution of applications to grind to a halt.  Under-utilization decreases resource utilization and can cause unnecessary expenditures in operating costs of a cluster. Thus, to maximize performance and resource utilization, an intelligent scheduler must take into account resource availability in the cluster as well as resource demand/requirement of a storm application in order to calculate an efficient scheduling. 

In this paper, we present the design of \textbf{\textit{R-Storm}} that implements resource-aware scheduling in Storm.  R-Storm significantly outperforms the default Storm as demonstrated in our evaluation.  The paper is organized in the following manner:  In Section \ref{overview}, we provide an overview of Storm. In Section \ref{problem}, we discuss the problem and provide the problem definition and formulation. Later in Section \ref{proposed}, we describe in detail the algorithms R-Storm uses for its resource-aware scheduling in Storm. In Section \ref{implementation}, we provide an overview on the architecture and implementation of R-Storm as well as a description of the User API we have implemented.  In Section \ref{evaluation}, we provide an evaluation of R-Storm. In Section \ref{relatedwork}, we discuss the body of related research work, and Section \ref{conclusion} includes our concluding comments.

Our work makes the following contributions: 1) We have created a system called R-Storm that, to the best of our knowledge, is the first system to implement resource-aware scheduling within Storm. R-Storm is able to support both hard and soft resource constraints. 2) We evaluate the performance of R-Storm on a range of micro-benchmarks as well as applications used in industry to demonstrate that R-Storm outperforms the default Storm in both overall throughput as well as resource utilization.  We also demonstrate that R-Storm is able to efficiently schedule multiple topologies. 

\section{Background}
\label{overview}
\begin{figure}[t!]
\centering
\includegraphics[width=\columnwidth]{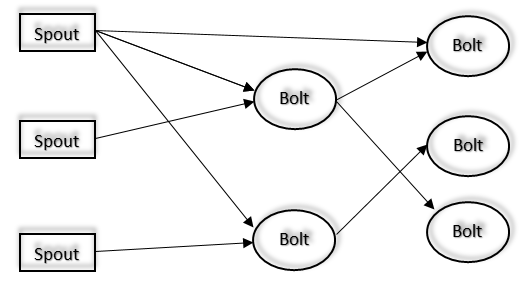}
\caption{An Example of Storm topology}
\label{StormTopology}
\end{figure}

Storm is a distributed processing framework that can process incoming live data in real-time \cite{storm_web}. Storm processes real-time data via ``topologies". A Storm \textit{topology} is a computation graph that provides a logic view of the data flow and how the data is processed. Similar to a Map-Reduce job, Storm jobs typically fragment the input dataset into independent chunks which are processed by the tasks. One major difference is that a MapReduce job finishes ultimately, while in Storm, a topology runs forever until it is killed. Figure~\ref{StormTopology} provides an example of a Storm topology. We define a list of terms used in Storm:
\begin{itemize}
\item
\textit{Tuples} - The basic unit of data that is processed.

\item
\textit{Stream} - an unbounded sequence of tuples.

\item
\textit{Component} - A processing operator in a Storm topology that is either a Bolt or Spout (defined later in the paper)

\item
\textit{Tasks} - A Storm job that is an instantiation of a Spout or Bolt (defined later in the paper).

\item 
\textit{Executors} - A thread that is spawned in a worker process (defined later) that may execute one or more tasks.

\textit{Worker Process} - A process spawned by Storm that may run one or more executors.

\end{itemize}

An processing vertex or operator in Storm is called a \textit{component}. A Storm topology contains two basic components: Spouts and Bolts.
\begin{enumerate}
\item
\textit{Spout} - This type of component is a source of data streams and it emits an unbounded number of tuples further downstream in the topology.   For example, a spout can be programmed to receive or read live financial data from the stock market and transform that data into tuples and emit them to be processed in some meaningful way further down in the Storm topology.

\item
\textit{Bolt} -  This is a component that consumes, processes, and potentially emits new streams of data.  A bolt can consume any number of input streams from spouts or other bolts, do some processing on the received data, and potentially emit new streams to be received and processed downstream. Bolts can filter tuples, perform aggregations, carry out joins, query databases, and in general, any user-defined functions. Multiple bolts can work together to compute complex stream transformations that may require multiple steps, like computing a stream of trending topics in tweets from Twitter \cite{storm_web}.
\end{enumerate}

A Storm Cluster has two types of nodes:

\begin{enumerate}

\item
\textit{Master Node} – This is the node responsible for scheduling tasks among worker nodes and also maintains an active membership list to ensure reliable fault-tolerant processing of data. The master node runs a daemon called “Nimbus”.  Nimbus communicates and coordinates with Zookeeper\cite{zookeeper_web} to maintain a consistent list of active worker nodes and to detect failure in the membership. 
\item

\textit{Worker Node} - Most machines in a Storm cluster are worker nodes. Each worker node runs a daemon called the “Supervisor”.  The supervisor continually listens for the master node to assign it tasks to execute.  Each worker contains many worker processes which are the actual containers for tasks to be executed.  A worker node can have any number of worker processes executing on it to facilitate multiplexing.
\end{enumerate}

\begin{figure}[t!]
\centering
\includegraphics[width=\columnwidth]{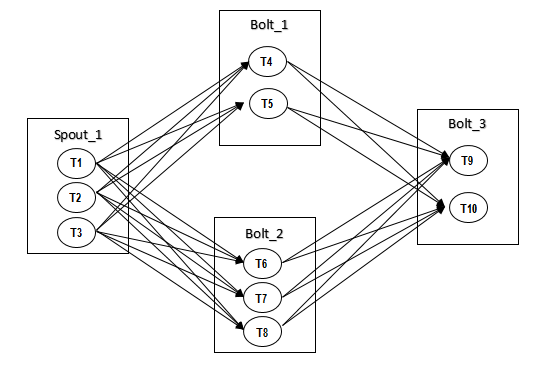}
\caption{Intercommunication of tasks within a Storm topology}
\label{StormComponents}
\end{figure}

\begin{figure}[t!]
\centering
      \includegraphics[width=\columnwidth]{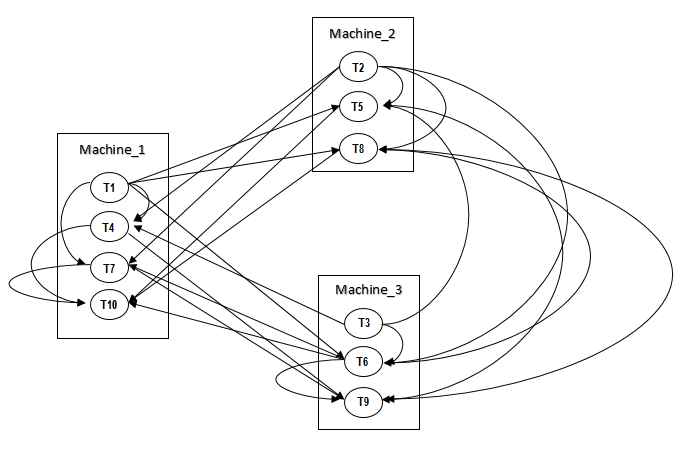}
  \caption{An example Storm machine}
 \label{StormMachine}
\end{figure}


Links between components in a Storm topology indicate how tuples are passed around. For example, if there is a directed link from a component A to a component B, that means component A will send to component B a stream of tuples. Each component in a Storm topology can be parallelized to potentially improve throughput.  The user needs to explicitly specify a parallelization hint for each component to set how many concurrent tasks to run for that component.  Each of these concurrent tasks that is parallelized from a component contains the same processing logic but may be executed at different physical locations and receive data from different sources.  Figure \ref{StormComponents} provides an example.

Storm's default Storm scheduler, which is a part of Nimbus daemon on the master node, will place tasks of bolts and spouts on worker nodes running worker processes in a round robin manner so tasks from a single bolt or spout will most likely be placed on different physical machines.  Figure~\ref{StormComponents} shows the intercommunication of tasks organized by component, however the actual intercommunication relative to the physical network and machines can be quite different. Figure~\ref{StormMachine} shows a example of a Storm cluster of three machines running the topology shown in Figure~\ref{StormComponents}. In Figure~\ref{StormMachine}, tasks are scheduled in a round robin fashion across all available machines.
\section{Problem Definition}
\label{problem}

The gist of problem we are trying to solve is how to best assign tasks to machines.  Each task has a set of certain resource requirements and each machine in a cluster has a set of resources that are available. Given these resource requirements and resource availability how can we create a scheduling such that, for all tasks, every resource requirement is satisfied if possible. Thus, the problem becomes find a mapping of tasks to machines such that every resource requirement is satisfied and at the same time no machine is exceeding its resource availability.

In this work, we consider three different types of resources: CPU usage, memory usage, and bandwidth usage. We define CPU resources as the projected CPU usage or availability percentage, memory resources as the megabytes used or available, and bandwidth as the network distance between two nodes.  We classify them as two different classes: \textit{hard} constraints and \textit{soft} constraints. Resources labeled as hard constraints must be satisfied to its full amount, while resources labeled as soft constraints may not be completely satisfied, however, we aim to minimize the number and amount of soft constraints that are violated. In the context of our system, CPU and bandwidth budgets are considered soft constraints as they can be overloaded, while the memory is considered as a hard constraint as we cannot exceed the total amount of available memory on a machine.  

In general, the number of constraints to use and whether a constraint is soft or hard is specified by the user.  The reasoning behind having hard and soft constraints is that some resources have a graceful degradation of performance while others do not.  For example, the performance of computation and network resources degrade as over utilization increases. However, if a system attempts to use more memory resources than physically available the consequences are catastrophic.  We have also identified that to improve performance sometimes it is beneficial to over utilize one set of soft-constrained resources but gain better utilization of another set of soft-constrained resources.  We assume that for each node of the cluster there is a specific limited budget for these resources. Similarly, each task specifies how much of each type of resource it needs.  Thus, this problem can essentially be modeled as a linear programming optimization problem as we discuss next. 

Let $\mathcal{T} = (\tau_1,\tau_2,\tau_3,\ldots)$ be the set of all tasks within a topology. Each task $\tau_i$ has a soft constraint CPU requirement of $c_{\tau_i}$, a soft constraint bandwidth requirement of $b_{\tau_i}$, and a hard constraint memory requirement of $m_{\tau_i}$. We discuss the notion of soft and hard constraints more in the next section. Similarly, let $\mathcal{N} = (\theta_1,\theta_2,\theta_3,\ldots)$ be the set of all nodes, which correspond to total available budgets of $W_1$, $W_2$, and $W_3$ for CPU, bandwidth, and memory. For the purpose of quantitative modeling, let the throughput contribution of every sink component to be $Q_{\theta_i}$ defined as the rate of tuples being processed at that node. The goal is to assign tasks to a subset of nodes $\mathcal{N'}\subseteq \mathcal{N}$ that increases the total throughput by maximizing resource utilization and minimizing network latency while at the same time not creating schedulings that will exceed the budgets $W_1$, $W_2$, and $W_3$. In other words

\begin{equation}
\begin{split} 
\textrm{Maximize }_{}\sum _{i \in {clusters}}~\sum _{j \in {nodes}}Q_{\theta_{ij}}
\\ \textit{  }  \textrm{ s.t. }\sum _{i \in {clusters}}~\sum _{j \in {nodes}} c_{\tau_{ij}} \leq W_1,\\
and \sum _{i \in {clusters}}~\sum _{j \in {nodes}} b_{\tau_{ij}} \leq W_2,\\
and \sum _{i \in {clusters}}~\sum _{j \in {nodes}} m_{\tau_{ij}} \leq W_3.
\end{split}
\end{equation}
which is simplified into

\begin{equation}
\begin{split} 
\textrm{Maximize }_{\{\mathcal{N'} \subseteq \mathcal{N}\}}\sum _{\theta \in \mathcal{N'}}Q_{\theta_i} \textit{  }  \textrm{   subject to  } \\
\sum_{\tau_i \in \mathcal{N'}} c_{\tau_i} \leq W_1, 
\sum_{\tau_i \in \mathcal{N'}} b_{\tau_i} \leq W_2,
\sum_{\tau_i \in \mathcal{N'}} m_{\tau_i} \leq W_3.
\end{split} 
\label{eq.1}
\end{equation}

This selection and assignment scheme is a complex and a special variation of Knapsack optimization problem. The well-known binary Knapsack problem is NP-hard but efficient approximation algorithms can be utilized (fully polynomial approximation schemes), so an approach is computationally feasible. However, the binary version, which is the most common Knapsack problem only considers a single constraint and enables only a \textit{subset} of the tasks to be selected, which is not desired as we need to assign \textit{all} the tasks to the nodes, considering that there are multiple constraints. To overcome the shortcomings of the binary Knapsack approach, we need to formulate the problem as other variations of the Knapsack problem, that eventually assigns \textit{all} the tasks to nodes. In our problem formulation we identify three challenges.

The first challenge is the fact that our problem consists of multiple knapsacks (i.e. clusters and corresponding nodes). This may seem like a trivial change, but it is not equivalent to adding to the capacity of the initial knapsack. This variation is used in many loading and scheduling problems in Operations Research
\cite{Chekuri:2000:PMK:338219.338254}. Thus our problem corresponds to a \textit{Multiple Knapsack Problem} (MKP), in which we assign tasks to multiple different constrained nodes.

The second challenge that we need to consider in our formulation is that if there is more than one constraint for each knapsack (for example, given knapsack scenario, both a volume limit and a weight limit, where the volume and weight of each item are not related), we get the multidimensional knapsack problem, or \textit{m-Dimensional Knapsack Problem}. In our problem, we need to address 3 different resources (i.e. CPU, bandwidth, and memory constraints) which leads to a 3-Dimensional Knapsack Problem. 

The third challenge that we need to consider is that given the topology, assigning two successive tasks on the same node is more efficient than assigning them on two different nodes, or even two different clusters. This is another special variation of knapsack problem, called Quadratic Knapsack Problem (QKP) introduced by Gallo \textit{et al.}\cite{gallo1}\cite{gallo2}, which consists in choosing elements from \textit{n} items for maximizing a quadratic profit objective function subject to a linear capacity constraint.

Our problem is a \textit{Quadratic Multiple 3-Dimensional Knapsack Problem} (we call it QM3DKP). 
Different variations of the knapsack problem has been applied to certain contexts. Y. Song \textit{et al.} in their paper \cite{4753629} investigated the multiple knapsack problem and its applications in cognitive radio networks. Hosseini \textit{et al.} \cite{mmsys13,movid14,mmve15} applied the concept of multiple-choice knapsack problem to the context of multimedia streaming applications to save bandwidth and energy.
X. Xie and J. Liu in their paper\cite{4223174} studied QKP, while also in\cite{6848188}, the authors applied the concept of m-dimensional knapsack problem to the packet-level scheduling problem for a network, and proposed an approximation algorithm for that.

Many algorithms have been developed to solve various knapsack problems using dynamic programming \cite{andonov2000unbounded} \cite{martello1999dynamic}, tree search (such as A*) techniques used in AI \cite{greenberg1970branch} \cite{sarkar1991reducing}, approximation algorithms \cite{chekuri2005polynomial} \cite{fayard1995approximation}, and etc.  However, these algorithms will not necessarily produce a solution that overcomes all the challenges we laid out for our problem. Moreover, these algorithms are constraining in terms of computational complexity even though some algorithms have pseudo polynomial runtime complexity.  Most, if not all, of these algorithms would require much more time to compute a scheduling that than necessarily available in a distributed data stream system.  Since data stream systems like Storm are required to respond to events as close to real-time as possible, scheduling decisions need to be made in a snappy manner.  The longer the scheduling takes to compute, the longer the downtime an application will have.  Moreover, if there are failures in the Storm cluster and executors need to be rescheduled, the scheduler must be able to produce another scheduling quickly.  If executors are not rescheduled quickly, whole topologies my be stalled, or worst, catastrophic and cascading failures might occur do to overflowing of message queues.

Thus, we need a scheduling algorithm that can schedule all tasks to multiple nodes and at the same time respect all resource requirements while with a high probability schedule two successive tasks on the same node. The algorithm needs to be simple with low overhead for it to be suitable for real-time requirements of Storm applications.  In the next section, we discuss the details of R-Storm and how it schedules tasks within Storm.
	
\section{R-Storm Scheduling Algorithm}
\label{proposed}
As discussed in the previous section, producing an optimal solution to our resource-aware scheduling problem in Storm can be very difficult and computationally infeasible. Therefore, we need simpler yet effective algorithms that circumvents the challenges involved in solving knapsack problems and more specifically the Quadratic Multiple 3-Dimensional Knapsack Problem in our case.

\begin{figure}[t]
	\centering
	\includegraphics[width=0.4\textwidth]{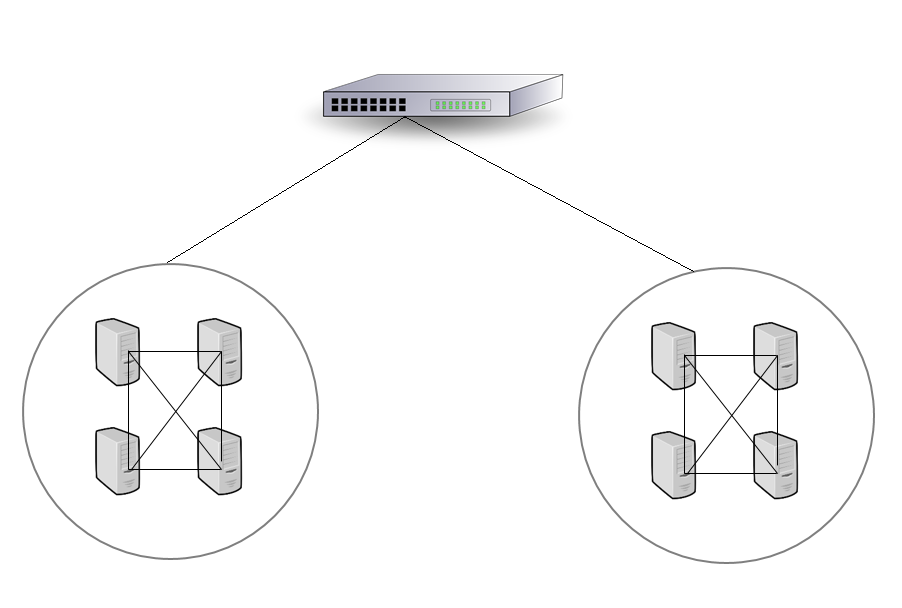}
	\caption{Typical cluster layout}
	\label{cluster_pic}
\end{figure}

In our pursuit of designing a scheduling algorithm, we made some important observations about the environment in which Storm is running in. Storm is usually deployed in data centers where servers are placed on a rack connected to each other by a top-of-rack switch. The top-of-rack switch is also connected to another switch which connects server racks together. The network layout can be represented by Figure~\ref{cluster_pic}. We also gained some insight into how communication latency is affected by the network distance.

To consider intercommunication demands for our formulation, we designed our scheduling algorithm around the insight:
\begin{enumerate}
\item
Inter-rack communication is the slowest
\item
Inter-node communication is slow
\item
Inter-process communication is faster
\item 
Intra-process communication is the fastest
\end{enumerate}

As discussed in the previous section, let $\mathcal{T} = (\tau_1,\tau_2,\tau_3,\ldots)$ be a topology. A task has a set of resource requirement needed in order to execute. Thus, a task $\tau_i$ has two soft resource constraints $c_{\tau_i}$ and $b_{\tau_i}$ and a hard resource constraint $m_{\tau_i}$. A cluster can be represented as a set of nodes $\mathcal{N} = (\theta_1,\theta_2,\theta_3,\ldots)$.  Every node has a specific available memory, CPU, and network resources. A node $\theta_i$ has a resource availability denoted by $c_{\theta_i}$,$b_{\theta_i}$, and $m_{\theta_i}$ for CPU, bandwidth, and memory, respectively.

Therefore, the resource demand for a task $\tau_i$ can be denoted as a set or 3-dimensional vector:
\\
\begin{center}
$A_{\tau_i} = \{m_{\tau_i},c_{\tau_i},b_{\tau_i}\}$
\end{center}

and a set of soft constraints:

\begin{center}
$S_{\tau_i} = \{c_{\tau_i},b_{\tau_i}\}$
\end{center}

and set of hard constraints:

\begin{center}
$H_{\tau_i} = \{m_{\tau_i}\}$
\end{center}

Such that
\begin{center}
$H_{\tau_i} \subseteq A_{\tau_i}$

$S_{\tau_i} \subseteq A_{\tau_i}$

\end{center}

and,

\begin{center}
$A_{\tau_i} = S_{\tau_i} \cup H_{\tau_i}$
\end{center}

The resource availability of a node $\theta_i$ can be denoted as a set or 3-dimensional vector:

\begin{center}
$A_{\theta_i} = \{m_{\theta_i},c_{\theta_i},b_{\theta_i}\}$
\end{center}

and a set of soft constraints:

\begin{center}
$S_{\theta_i} = \{c_{\theta_i},b_{\theta_i}\}$
\end{center}

and set of hard constraints:

\begin{center}
$H_{\theta_i} = \{m_{\theta_i}\}$
\end{center}

Such that

\begin{center}
$H_{\theta_i} \subseteq A_{\theta_i}$

$S_{\theta_i} \subseteq A_{\theta_i}$

\end{center}

and,

\begin{center}
$A_{\theta_i} = S_{\theta_i} \cup H_{\theta_i}$
\end{center}

This formulation can easily be generalized to model the resource availability of a node and the resource demand of a specific task as a $n$-dimensional vector residing in $\mathbb{R}^n$. Each soft constraint can have a weight attached to it, such that:

\begin{center}
$|Weights| = |S|$

$S^{'} = Weights \cdot S$

\end{center}

The reason for allowing constraints to be weighted is so that values can be normalized for comparison, as well as for allowing users to decide which constraints are more valued.

\subsection{Algorithm Overview}
\label{algorithm}
 
\begin{figure}[t!]
\centering
\includegraphics[width=\columnwidth]{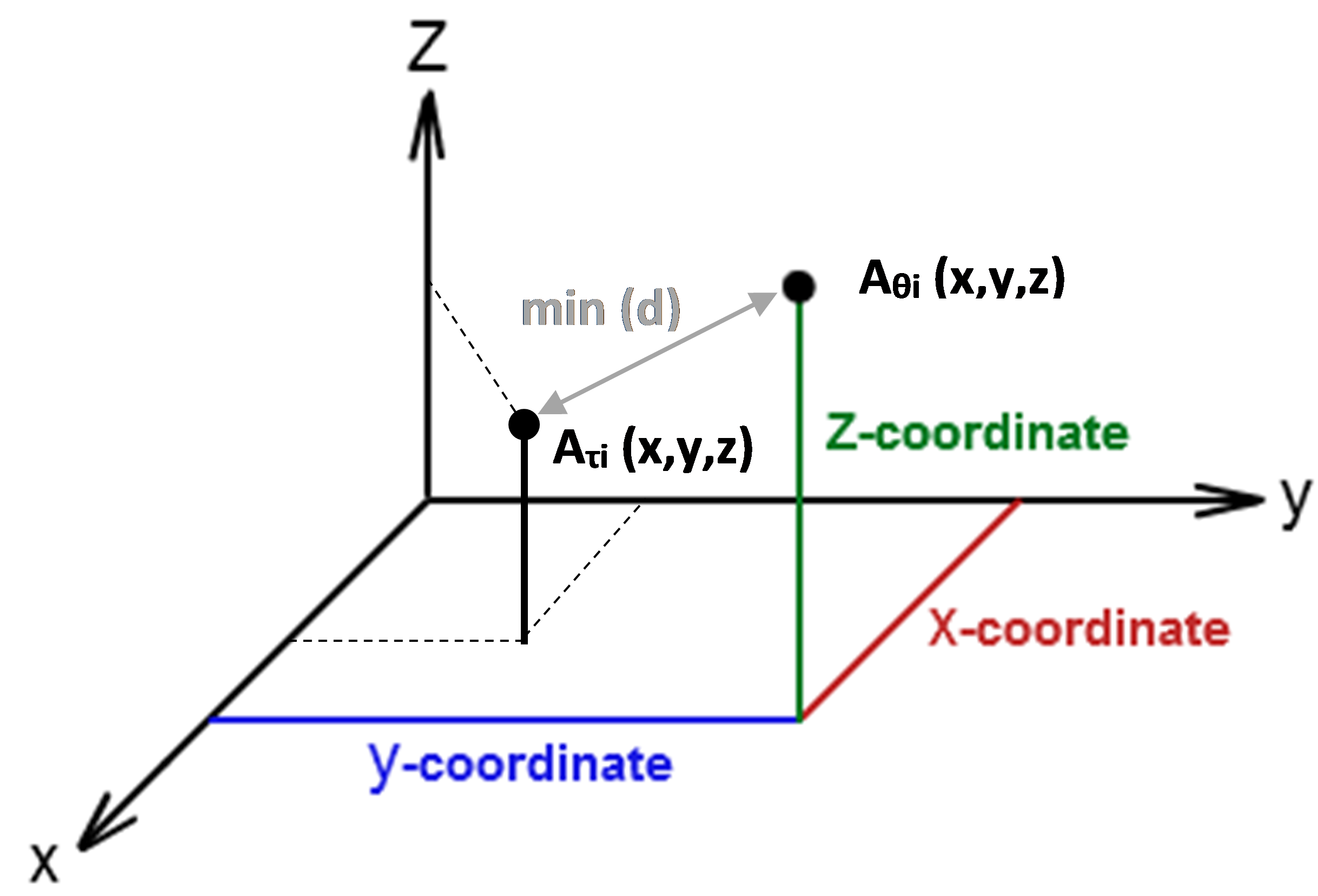}
\caption{An example node selection in a 3D resource space}
\label{3dr}
\end{figure}

Given a Storm topology  $\mathcal{T}$ consisting of a set of tasks $(\tau_1,\tau_2,\tau_3,\ldots)$ and a cluster $\mathcal{N}$ consisting of a set of nodes $(\theta_1,\theta_2,\theta_3,\ldots)$ where each node $\theta_i$ has a corresponding vector $A_{\theta_i}$ representing the node's resource availability.    For each task $\tau_i \epsilon \mathcal{T}$ and its corresponding resource demand vector $A_{\tau_i}$, the algorithm determines which node to schedule a task on.  Assuming that each resource corresponds to an axis, we find the node $A_{\theta_i}$ that is closest in Euclidean distance to $A_{\tau_i}$ such that $H_{\theta_i} > H_{\tau_i}$ for all hard constraints, so that no hard constraints are violated. We choose $\theta_i$ as the node where we schedule task $\tau_i$, and then update the available resources left on $A_{\theta_i}$. We continue this process until all tasks get scheduled on nodes. Our proposed heuristic algorithm has the following properties:
\begin{enumerate}
\item Tasks of components that communicate with each other will have the highest priority to be scheduled in close network proximity to each other.
\item No hard resource constraints is violated.
\item Resource wastes on nodes are minimized.
\end{enumerate}

Figure \ref{3dr} shows a visual example of a selected minimum-distance node to a given task in the 3D resource space, while the hard resource constraint (i.e. Z axis) is not violated.  

\begin{algorithm}[t]
 \caption{R-Storm Schedule}
 \label{algorithm:4}
\begin{algorithmic}[1]
\Procedure{Schedule}{}
	\State $taskOrdering \leftarrow$ \Call{TaskSelection}{()}
	\For{each Task $\tau$ in $taskOrdering$}
		\State Node $n \leftarrow$ \Call{NodeSelection}{$\tau$, $cluster$}
		\State\Call{schedule}{$\tau$, $n$};
	\EndFor
\EndProcedure
\end{algorithmic}
\end{algorithm}

Our algorithm consists of two core parts, task selection and node selection. Algorithm~\ref{algorithm:4} depicts the pseudo-code for scheduling procedure in R-Storm.  When scheduling, R-Storm first obtains a ordered list of tasks to be scheduled via $TaskSelection$ procedure (Algorithm~\ref{algorithm:4} line 2). Then, for each task in the ordered list, we find a node on which the task will run via the $NodeSelection$ procedure (Algorithm~\ref{algorithm:4} line 3-6). Please note that the actual assignment of task to node is done in an atomic fashion after the schedule mapping between all tasks to nodes has been determined. Next, we explain the process of task selection and node selection in detail.



\subsubsection{Task Selection}
\begin{algorithm}[t]
 \caption{Topology Traversal}
 \label{algorithm:2}
\begin{algorithmic}[1]
\Procedure{BFSTopologyTraversal} {Component $root$}
	\State $queue;$ \% queue of Components
	\State $visted;$ \% list of Components
	\If{$root$ == $null$}
	
		\Return{$null$}
		
	\EndIf
	\State $queue.add(root)$
	\State $visted.add(root)$
	\While{$queue$ is not empty}
		\State Component $com \leftarrow queue.remove()$ 
		\For{each Component $n$ in $com.neighbor$}
			\If{$visited$ does not contain $n$} \%check whether visited or not
				\State $queue.add(n$);
				\State $visited.add(n)$

			\EndIf
		\EndFor
	\EndWhile
	
	\Return{$visited$}
	
\EndProcedure
\end{algorithmic}
\end{algorithm}

\begin{algorithm}[t]
 \caption{Task Selection}
 \label{algorithm:3}
\begin{algorithmic}[1]
\Procedure{TaskSelection}{}
	\State $components \leftarrow$ \Call{BFSTopologyTraversal}{$root$}
	\While{$taskOrdering$ does not contain all tasks}
		\For{each Component $c$ in $components$ }
			\If{$c$ has tasks}
				\State Task $\tau \leftarrow c.getTask()$
				\State $taskOrdering.add(\tau)$
				\State $c.removeTask(\tau)$
			\EndIf
		\EndFor
	\EndWhile
	
	\Return{$taskOrdering$}
	
\EndProcedure
\end{algorithmic}
\end{algorithm}

\begin{algorithm}[t]
	\caption{Node Selection}
	\label{algorithm:1}
	\begin{algorithmic}[1]
		\Procedure{NodeSelection}{Task $\tau$, Cluster $cluster$}
		\State $\mathcal{A_{\theta}}$: set of all nodes $\theta_i$ in the n-dimensional space~~~ \%(n=3 in our context)
		\State $\mathcal{A_{\tau}}$: set of all tasks $\tau_i$ in the n-dimensional space~~~ \%(n=3 in our context)
		\State $A_{\theta_i}$: n-Dimensional vector of resource availability on each node $\theta_i$ , such that $A_{\theta_i} \in \mathcal{A_{\theta}}$
		\State $A_{\tau_i}$: n-Dimensional vector of resource availability for each task $\tau_i$ , such that $A_{\tau_i} \in \mathcal{A_{\tau}}$
		\If{global $refNode$ == null}
		\State ServerRack $s \leftarrow$ findServerRackWithMostResources($cluster$)
		\State $refNode \leftarrow$ findNodeWithMostResources($s$)
		\EndIf
		
		\State select $A_{\theta_j}$ such that:
		
		$A_{\theta_j} = min \quad d(A_{\tau_i}, A_{\theta_j})\ \forall A_{\theta_j} \ \epsilon \ A_\theta$ given $d(\lambda, \lambda^2) \leftarrow$ \Call{Distance}{$\tau_i$, $\theta_j$}, $H_{\theta_j} > H_{\tau_i}$ \%for all hard resource constraints
		
		\Return{$\theta_j$}
		\EndProcedure	
		
		\Procedure{distance}{Task $\tau_i$, Node $\theta_j$}
		\State $distance \leftarrow weight_m *(m_{\tau_i} - m_{\theta_j})^2+weight_c *(c_{\tau_i} - c_{\theta_j})^2+weight_b * (newtorkDistance(refNode, \theta_j))$
		
		\Return {$\sqrt{distance}$}
		\EndProcedure
		
	\end{algorithmic}
\end{algorithm}

\begin{figure*}
	\centering
	\includegraphics[width=0.8\textwidth]{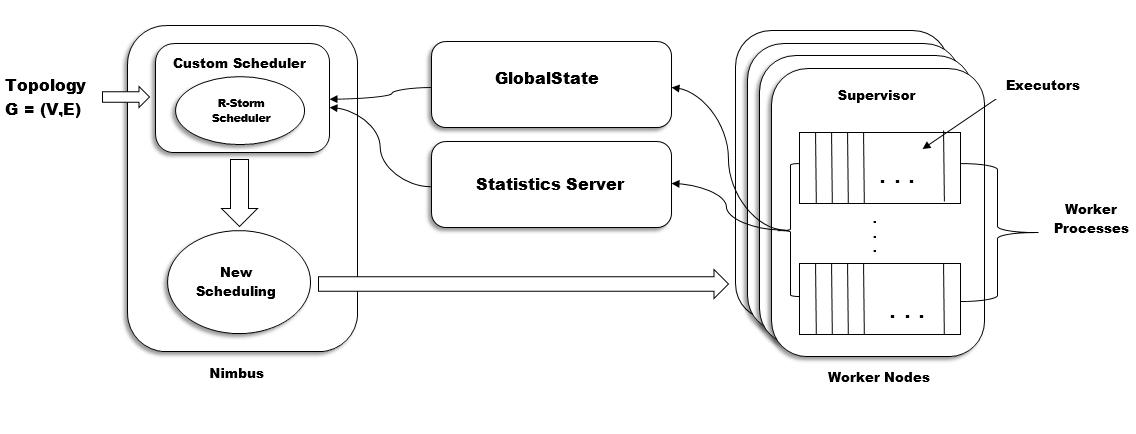}
	\caption{R-Storm Architecture Overview.}
	\label{architecture_diagram}
\end{figure*}

\label{taskselection}
In this section, we discuss in detail how we determine the ordering in which tasks are scheduled.   Algorithm~\ref{algorithm:3} depicts the pseudo-code for the whole task selection procedure. The first part of task selection is determining how to traverse the directed graph representing a Storm topology as well as where to start the traversal.  The location at which the traversal of the topology starts should be related to how important the set of components in that location is in respect to the rest of the topology. A number of heuristics can be used to determine the starting point, however, for simplicity, in our algorithm, we start traversing the topology starting from the spouts since the performance of spout(s) impacts the performance of the whole topology.  As for the traversal algorithms, we use breadth first search (BFS) traversal. Since BFS traverses one level at a time, we use BFS traversal to create an partial ordering of components in which adjacent components will be placed in close succession to each other. Algorithm~\ref{algorithm:2} depicts the pseudo-code for our BFS traversal of a Storm topology.

Once we have obtained a partial ordering of components from the BFS traversal, we create an partial ordering of tasks.  Lines 3-12 in Algorithm~\ref{algorithm:3} depicts this process. To create the ordering of tasks, we first iterate through ordered list of components we obtained from BFS traversal (Algorithm~\ref{algorithm:3} line 3).  For each component we iterated, we get \textit{one} task from this component and add it to our $taskOrdering$ list.  We keep iterating through the ordered list of components until we have added all tasks into $taskOrdering$. Ordering tasks to be scheduled in this fashion will ensure that tasks from adjacent components will be scheduled as close together as possible, thus fulfilling the first desired property of our scheduling algorithm that we listed in the previous section.
\subsection{Node Selection}
After we have obtained an ordered list of tasks to schedule, we need to determine on which node to schedule each task.  Algorithm~\ref{algorithm:1} depicts the pseudo-code for node selection.
After a task is selected to be scheduled, a node needs to be selected to run this task by invoking the $NodeSelection$ procedure in Algorithm~\ref{algorithm:1}.  If the task that needs to be scheduled is the first task in a topology, find the server rack or sub-cluster with the most available resources. Afterwards, find the node in that server rack with the most available resources and schedule the first task on that node which we refer to as the \textit{Ref Node} (Algorithm~\ref{algorithm:1} lines 6-9). For the rest of the tasks in the Storm topology, we find nodes to schedule based on the $Distance$ procedure in Algorithm~\ref{algorithm:1} with our bandwidth attribute $b_{\theta_i}$ defined as the network distance from \textit{Ref Node} to node $\theta_i$.  The $Distance$ procedure calculates the Euclidean distance between the resource requirement vector of a task $\tau_i$ and the resource availability vector of a node $\theta_j$.  By selecting nodes in this manner, tasks will be patched as tightly on or closely around the \textit{Ref Node} as resource constraints allow, which minimizes the network latency of tasks communicating with each other.  This process is visual depicted in Figure~\ref{3dr}.

\section{Implementation}
\label{implementation}
We have implemented R-Storm as a custom version of Storm. We have modified the core Storm code to allow physical machines to send their resource availability to Nimbus. The core scheduling functions of R-storm is implemented as a custom scheduler in our custom version of Storm. A user can create a custom scheduler by creating a Java class that implements a predefined IScheduler interface.  The scheduler runs as part of the Storm Nimbus daemon. A user specifies which scheduler to use in a YAML formated configuration file call \textit{storm.yaml}. The Storm scheduler is invoked by Nimbus periodically, with a default time period set to 10 seconds.  Storm Nimbus is a stateless entity and thus, any Storm scheduler does not have any mechanism to store any information across multiple invocations.  The architecture of R-Storm is graphically depicted in Figure~\ref{architecture_diagram}.
\subsection{Core Architecture}
Our implementation of R-Storm has three modules:
\begin{enumerate}
\item
StatisticServer - This module is responsible for collecting statistics in the Storm cluster, e.g., throughput on a task, component, and topology level. The data generated in this module is used for evaluative purposes.
\item
GlobalState - This module stores important state information regarding the scheduling and resource availability of a Storm Cluster. This module will hold information about where each task is placed in the cluster.  This module also stores all the resource availability information of physical machines in the cluster and the resource demand information of all tasks for all Storm topologies that are scheduled or needs to be scheduled.

\item
ResourceAwareScheduler - This module is the custom scheduler that implements IScheduler interface.  This class starts and initializes the StatisticServer and GlobalState modules.  This module also contains the implementation of the core R-Storm scheduling algorithm.
\end{enumerate}

\subsection{User API}
\begin{figure*}[t]%
\centering
\begin{subfigure}[t]{.65\columnwidth}
\centering
\includegraphics[width=\columnwidth]{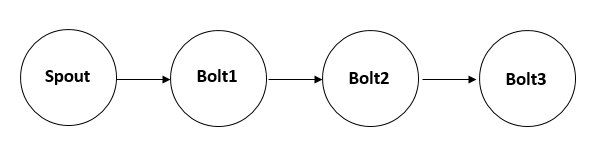}%
\caption{Layout of Linear Topology}%
\label{LinearTopologyLayout}%
\end{subfigure}\hfill%
\centering
\begin{subfigure}[t]{.65\columnwidth}
\includegraphics[width=\columnwidth]{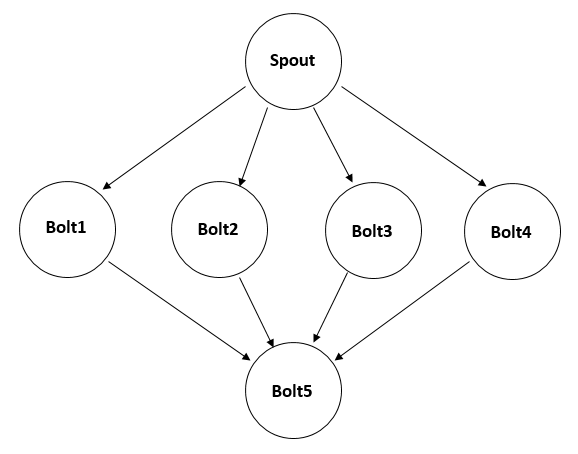}%
\caption{Layout of Diamond Topology}%
\label{DiamondTopologyLayout}%
\end{subfigure}\hfill%
\begin{subfigure}[t]{.65\columnwidth}
\centering
\includegraphics[width=\columnwidth,height=4cm, keepaspectratio]{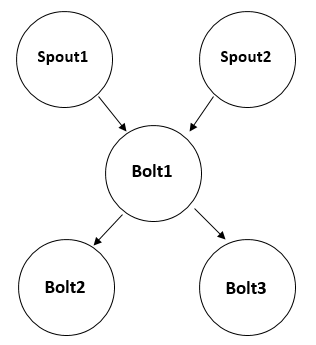}%
\caption{Layout of Star Topology}%
\label{StarTopologyLayout}%
\end{subfigure}%
\caption{Layout of Micro-benchmark Topologies}
\label{TopologyLayout}
\end{figure*}

\begin{figure*}[t]%
\centering
\begin{subfigure}{.65\columnwidth}
\includegraphics[width=\columnwidth]{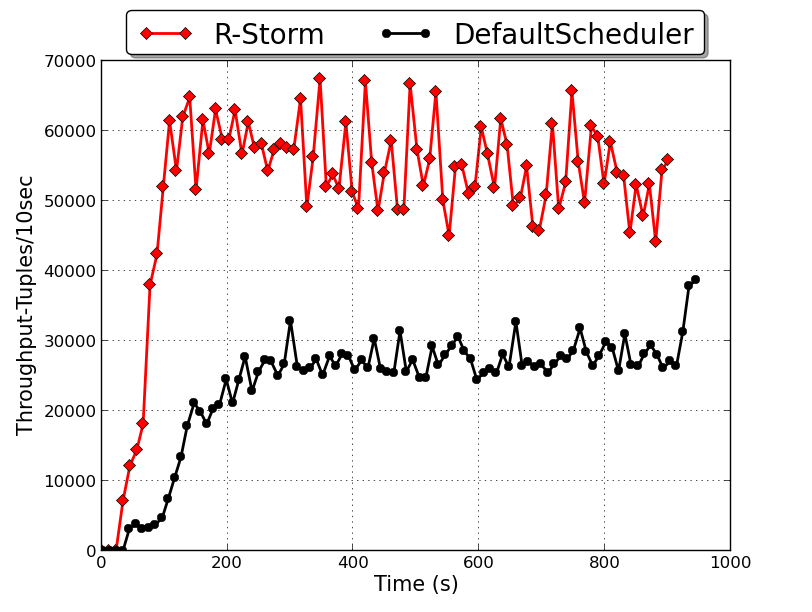}%
\caption{Linear Topology}%
\label{LinearTopologyResults}%
\end{subfigure}\hfill%
\begin{subfigure}{.65\columnwidth}
\includegraphics[width=\columnwidth]{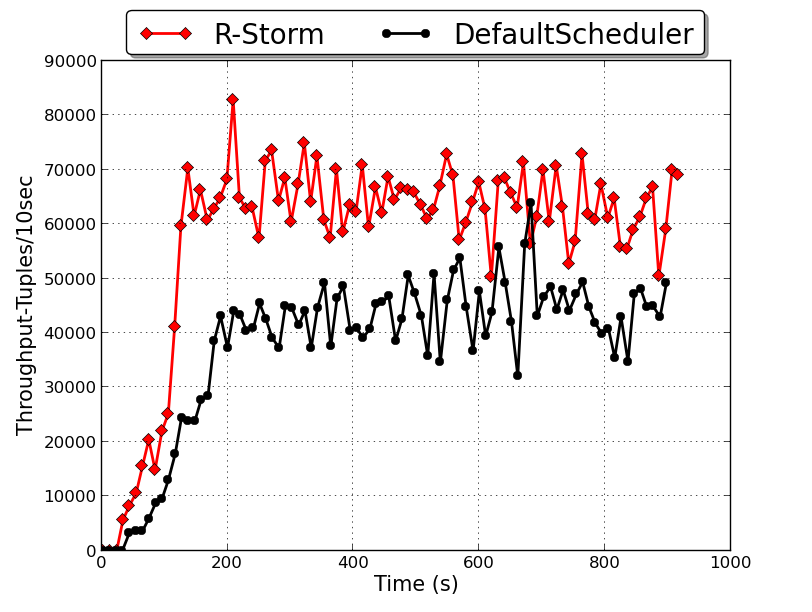}%
\caption{Diamond Topology}%
\label{DiamondTopologyResults}%
\end{subfigure}\hfill%
\begin{subfigure}{.65\columnwidth}
\includegraphics[width=\columnwidth]{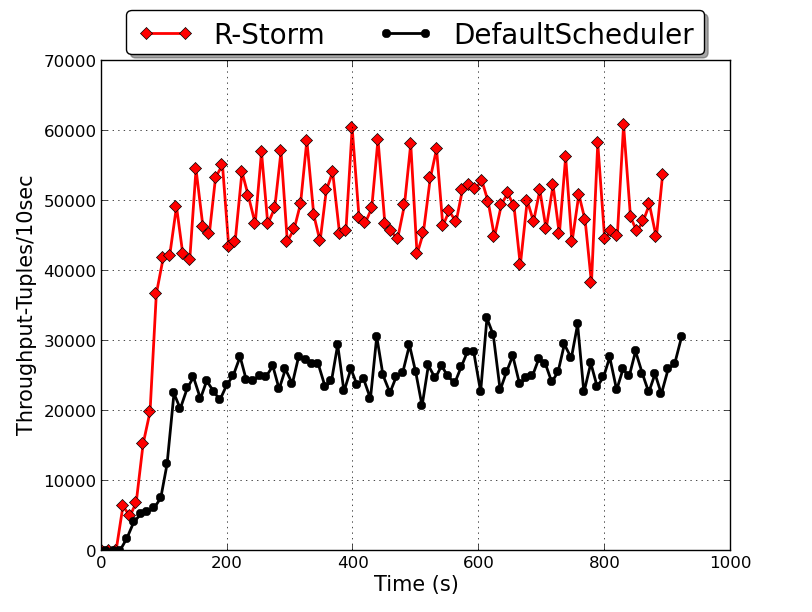}%
\caption{Star Topology}%
\label{StarTopologyResults}%
\end{subfigure}%
\caption{Experimental results of Network-bound Micro-benchmark Topologies}
\label{TopologyResults}
\end{figure*}
We have designed a list of APIs for the user to specify the resource demand of any component and the resource availability of any physical machine. For a Storm topology, the user can specify in the topology application code the amount of resources a topology component (i.e. Spout or Bolt) is required to run a single instance of the component by using the following API calls.
\\\\
\texttt{public T setMemoryLoad(Double amount)}
\\
Parameters:
\begin{itemize}
\item
\texttt{Double amount} – The amount of on memory an instance of this component will consume in megabytes.
\end{itemize}
~\\
\noindent
\texttt{public T setCPULoad(Double amount)}
\\
Parameters:
\begin{itemize}
\item
Double amount – The amount of on CPU an instance of this component will consume.
\end{itemize}
\noindent
Example of Usage:
\\\\
\texttt{SpoutDeclarer s1 = builder.setSpout("word", new TestWordSpout(), 10);}
\\\\
\texttt{s1.setMemoryLoad(1024.0);}
\\\\
\texttt{s1.setCPULoad(50.0);}
\\\\
Next, we discuss how to specify the amount of resources available on a machine.
An administrator can specify a machine's resource availability by modifying the \textit{conf/storm.yaml} file located in the storm home directory of that machine.

A administrator can specify how much available memory a machine has in megabytes by adding the following to \textit{storm.yaml}
\\\\
\texttt{supervisor.memory.capacity.mb:~[amount<Double>]}
\\\\
A administrator can specify how much available CPU a machine has adding the following to \textit{storm.yaml}
\\\\
\texttt{supervisor.cpu.capacity: [amount<Double>]}
\\\\
Example of usage:
\\\\
\texttt{supervisor.memory.capacity.mb: 20480.0}
\\\\
\texttt{supervisor.cpu.capacity: 100.0}
\\\\
Currently, the amount of CPU resources a component requires or is available on a machine is represented by point system since CPU usage is a difficult concept to define. Different CPU architectures perform differently depending on the task at hand. The point system is a rough estimate of what percentage of CPU core a task is going to consume.  Thus, for a typical situation, the CPU availability of a node is set to $100\ *\ \#\ of\ cores$.  For the purposes of this paper, we are assuming that the Storm cluster is homogeneous.

\section{Evaluation}
\label{evaluation}

In this section, we discuss how we evaluated the performance of R-Storm. We first present an overview of our experimental setup, followed by a comparison of performance of the R-Storm versus the default Storm measured on a variety of micro-benchmark Storm topologies.  Next, we compare the performance of R-Storm with default Storm on two Storm topologies used in production at Yahoo! Inc. We also present an evaluation of R-Storm scheduling multiple topologies.

 \begin{figure*}[t]%
 \centering
 \begin{subfigure}{.7\columnwidth}
 \includegraphics[width=\columnwidth]{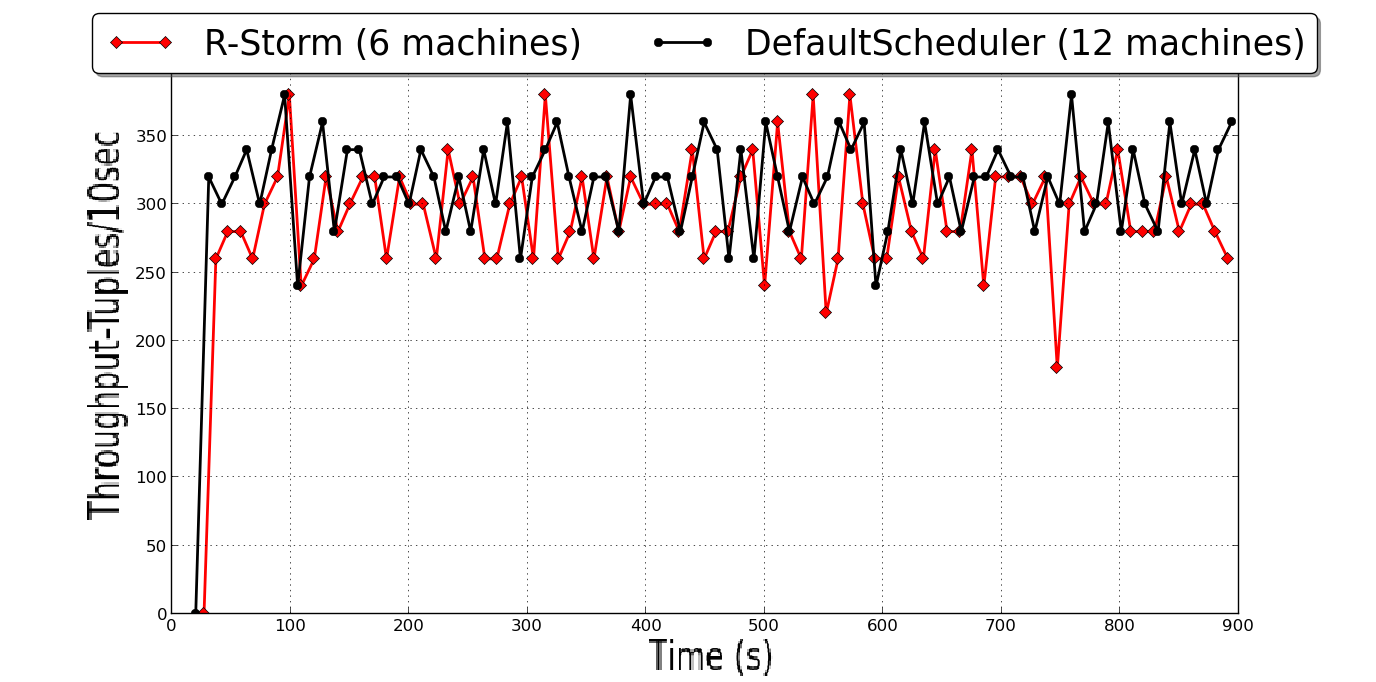}%
 \caption{Linear Topology}%
 \label{linearTopology_cpu_bound_throughput}%
 \end{subfigure}\hfill%
 \begin{subfigure}{.7\columnwidth}
 \includegraphics[width=\columnwidth]{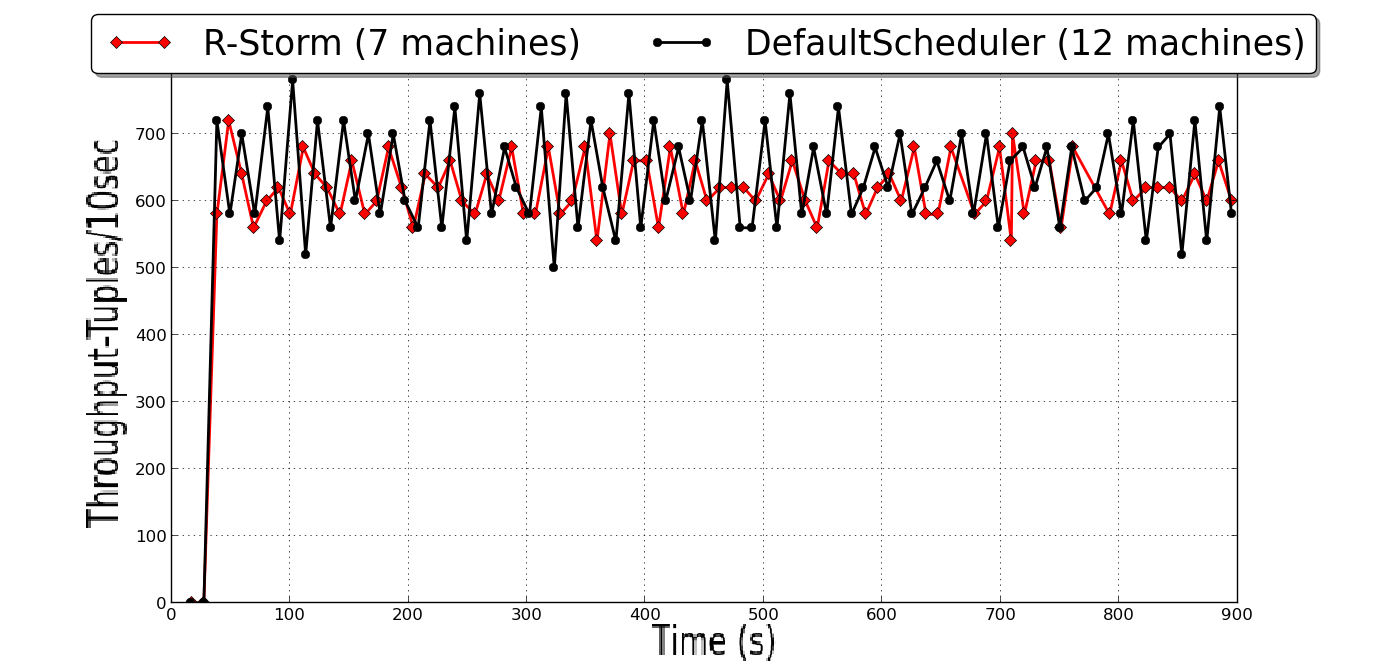}%
 \caption{Diamond Topology}%
 \label{diamond_cpu_bound_throughput}%
 \end{subfigure}\hfill%
 \begin{subfigure}{.7\columnwidth}
 \includegraphics[width=\columnwidth]{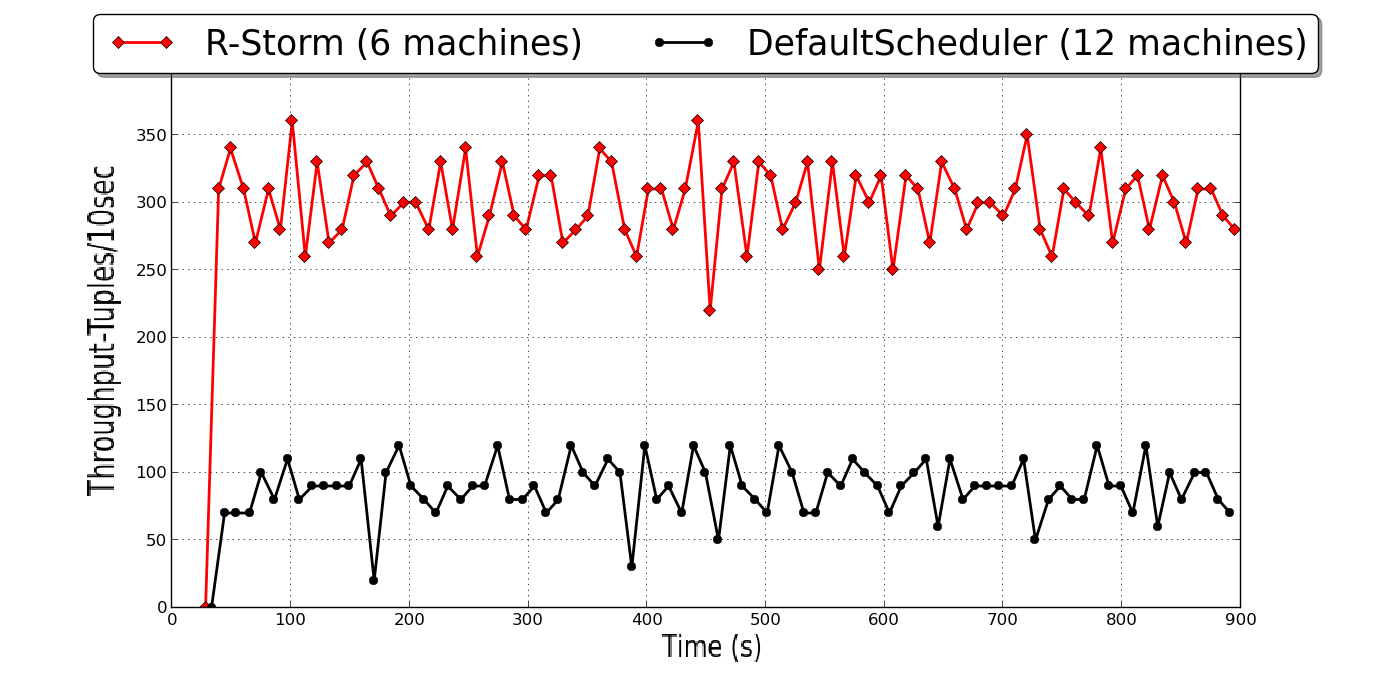}%
 \caption{Star Topology}%
 \label{star_cpu_bound_throughput}%
 \end{subfigure}%
 \caption{Experimental results of Computation-time-bound Micro-benchmark topologies}
 \label{cpu_bound_results}
 \end{figure*}

\subsection{Experimental Setup}
We aim to evaluate our scheduling algorithm in a simulated real world environment where a Storm cluster is composed of machines from more than one server rack. We used Emulab\cite{emulab_web} to run our experiments. Emulab is a network testbed which provides researchers with a wide range of environment to develop, debug, and evaluate their experimental system.

In our Emulab experimental setup, the Storm cluster consists of a total of 13 machines. One machine is designated as the master node, i.e. runs Storm Nimbus and Zookeeper, while the other 12 machines are worker nodes. To emulate the latency of inter-rack communication, we create two VLANs, with each VLAN holding 6 machines. The latency cost of the inter-rack communication is 4ms for a round trip time. Each machine runs on Ubuntu 12.04 LTS with a single 3GHz processor, 2GB of RAM, 15K RPM 146GB SCSI disks, and is connected via 100Mbps network interface cards.

\subsection{Experiment Results}

We evaluate the performance of R-Storm by comparing the throughput of a variety of topologies  scheduled by R-Storm against Storm's default scheduler. To be comprehensive in our evaluation, we conduct our experiments using both common micro-benchmark Storm topologies, as well as, actual Storm topologies deployed in industry. For our evaluation, the throughput of a topology is the average throughput of all output bolts which tends to be farthest downstream components in a Storm topology. Our experiments are run for an average of 15 minutes by which the throughput of the Storm Topology that is being evaluated should have stabilized and converged.

\subsection{Micro-benchmark Storm Topologies: Linear, Diamond, and Star}

To fairly evaluate R-Storm, we created Micro-benchmark topologies representing commonly found topologies, namely \textit{Linear Topology}, \textit{Diamond Topology}, and \textit{Star Topology}. These Storm Topologies are visually depicted in Figures~\ref{LinearTopologyLayout}, \ref{DiamondTopologyLayout}, and \ref{StarTopologyLayout}.  The Linear Topology is similar to the evaluation topology used in a related work on Storm scheduling \cite{Polo:2011:RAS:2188461.2188475}.

Among the three resources (i.e. memory, CPU, network) we consider for our evaluation, the performance of a topology is majorly influenced by CPU and network resources.  Since a Storm topology executes as fast as it can, the performance of the topology is likely to be bounded by either the maximum performance of the CPU or network.  Thus, Storm topologies can be classified into categories: 1) topologies bounded by network resources and 2) topologies bounded by computation time.  When a workload is bound by network resource usage,  the overall throughput is limited by the amount network bandwidth and latency. On the other hand, for a workload that is bounded by computation time, the overall throughput is limited by the processing time of each tuple.

We first present the results of the micro-benchmark topologies that are configured to be network resource bound and then we present the results of the micro-benchmark topologies that are configured to be computation time bound.
 \begin{figure}[t]
 	\centering
 	\includegraphics[width=0.5\textwidth]{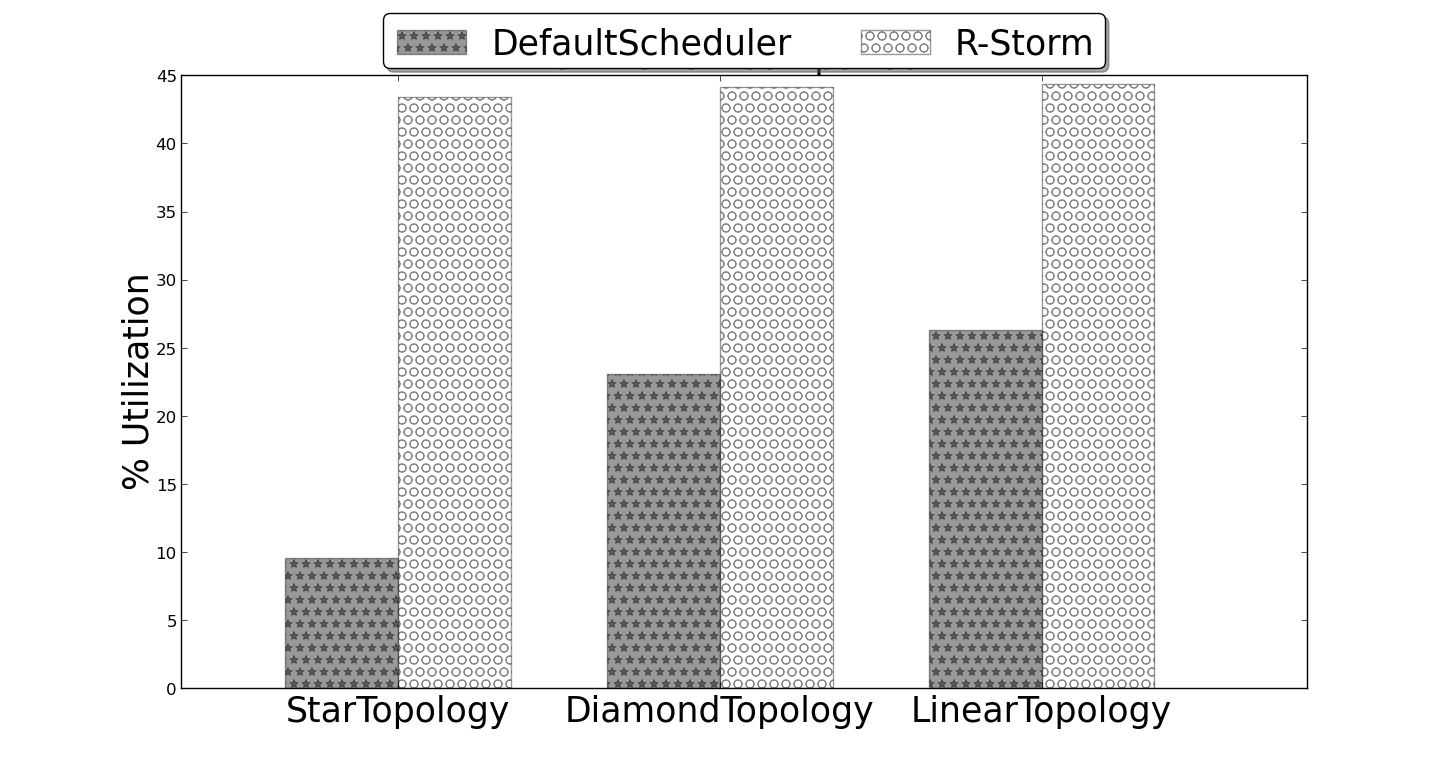}
 	\caption{CPU Utilization Comparison}
 	\label{cpu_util_compare}
 \end{figure}
\subsubsection{Network Resource Bound}
\begin{figure*}[t]%
	\centering
	\begin{subfigure}{\columnwidth}
		\includegraphics[width=\columnwidth]{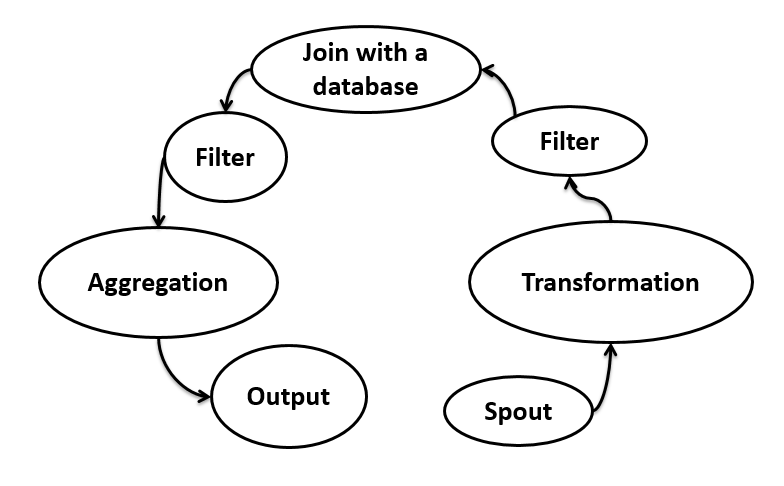}%
		\caption{Layout of Page Load Topology}%
		\label{PageLoadTopologyLayout}%
	\end{subfigure}\hfill%
	\begin{subfigure}{\columnwidth}
		\includegraphics[width=\columnwidth]{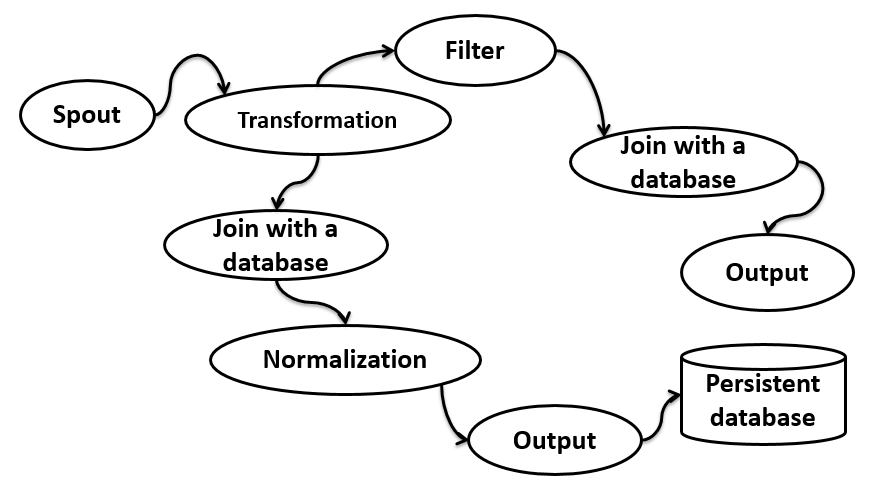}%
		\caption{Layout of Processing Topology}%
		\label{ProcessingTopologyLayout}%
	\end{subfigure}\hfill%
	\caption{Production Topologies Modeled After Typical Industry Topologies}
	\label{YahooTopologies}
\end{figure*}

\begin{figure*}[t]%
	\centering
	\begin{subfigure}{\columnwidth}
		\includegraphics[width=\columnwidth]{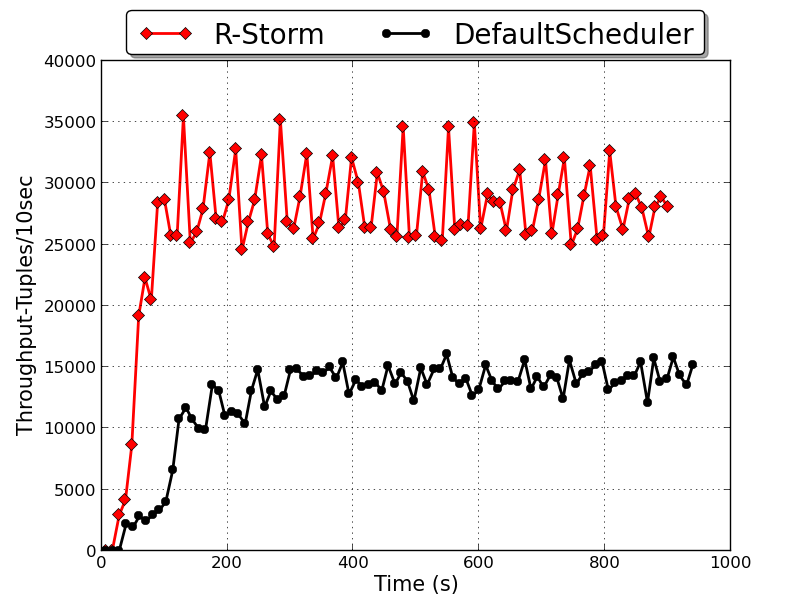}%
		\caption{Experiment results of Page Load Topology}%
		\label{PageLoadTopologyResults}%
	\end{subfigure}\hfill%
	\begin{subfigure}{\columnwidth}
		\includegraphics[width=\columnwidth]{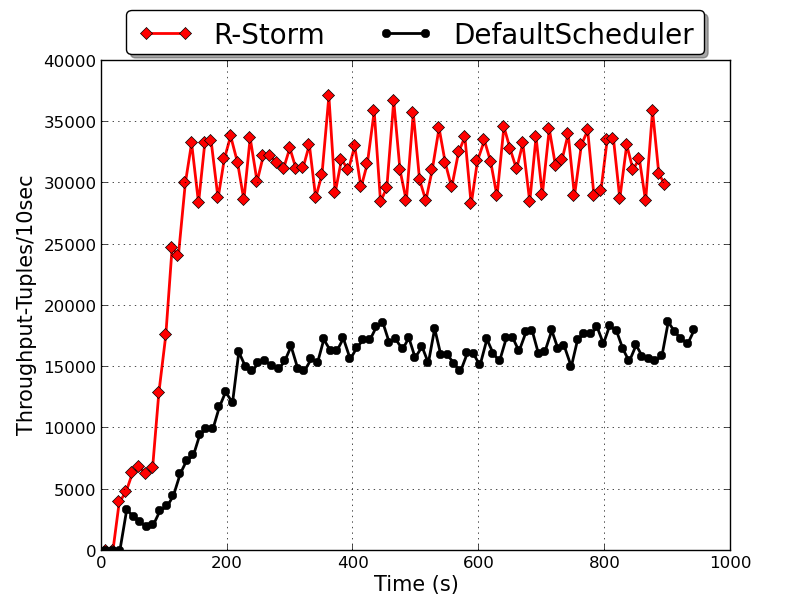}%
		\caption{Experiment results of Processing Topology}%
		\label{ProcessingTopologyResults}%
	\end{subfigure}\hfill%
	\caption{Experiment results of Industry Topologies}
	\label{YahooTopologiesResults}
\end{figure*}

In this scenario, we have configured the micro-benchmark topologies to do very little processing at each component.  Thus, the overall throughput is limited by the network speed in which tuples get sent through the network.  Figure~\ref{LinearTopologyResults}, \ref{DiamondTopologyResults}, and \ref{StarTopologyResults} compare the throughput of schedulings done by R-Storm versus that of Storm's default scheduler for the Linear, Diamond, and Star Topologies, respectively. The performance of those Storm topologies scheduled by R-Storm is significantly higher compared to the schedulings done by Storm's default scheduler. As seen in the results, scheduling computed by R-Storm provides on average of around 50\%, 30\%, and 47\% higher throughput than that computed by Storm's default scheduler, for the Linear, Diamond, and Star Topologies, respectively.  The significant improvement in throughput when using R-Storm can be attributed to R-Storm's ability to minimize network communication latency by colocating tasks that communicate with each other to the same machine or same server rack.


\begin{figure*}[t]
	\centering
	\includegraphics[width=0.8\textwidth, height=6cm]{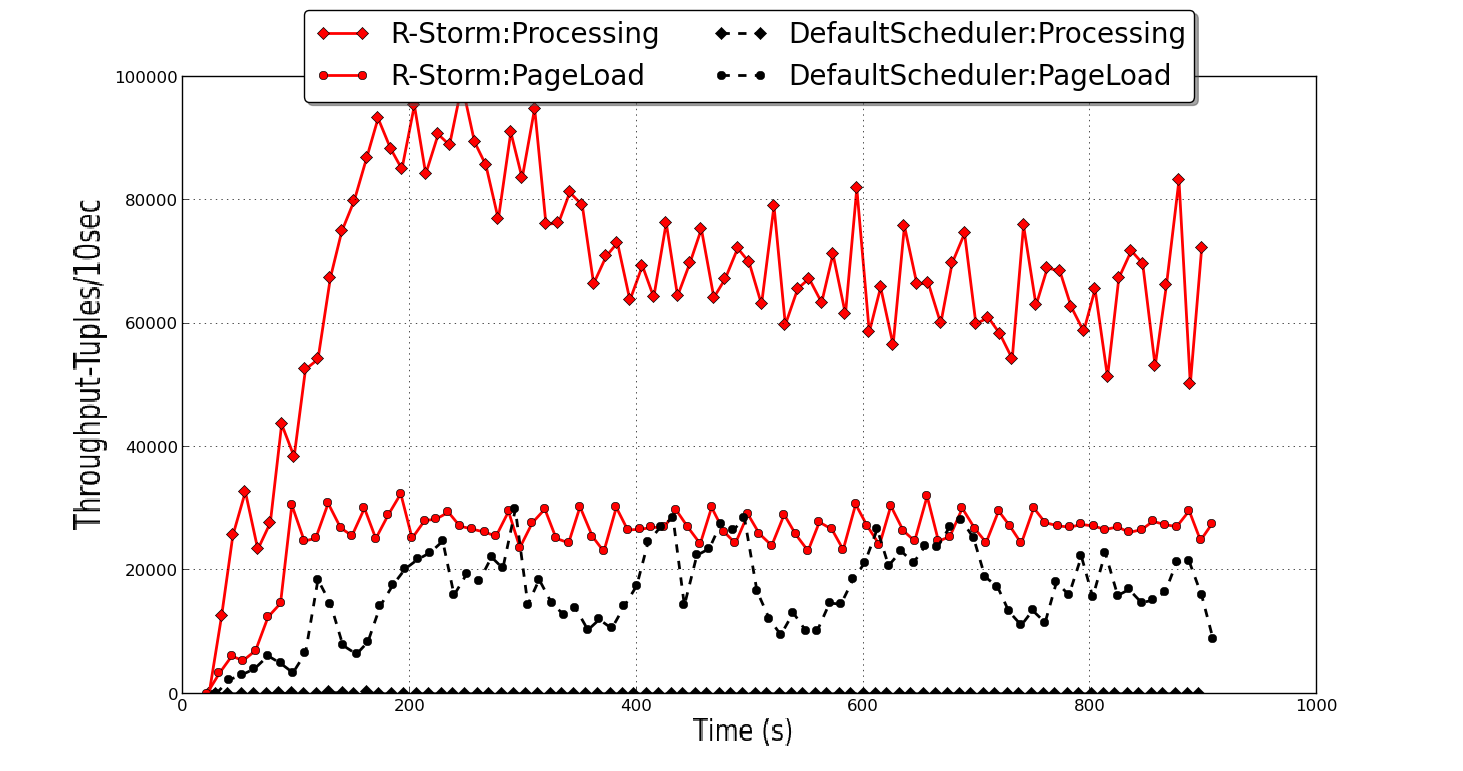}
	\caption{Throughput comparison of running multiple topologies.}
	\label{muli-topology-throughput}
\end{figure*}

\subsubsection{Computation Time Bound}

In this scenario, the overall throughput of the micro - benchmark topologies are bounded by the computation time spent at each component.  We have configured each component for the micro-benchmark topologies to conduct a significant amount of arbitrary processing. In this experiment, we intend to show that R-Storm can efficiently schedules tasks closely together to maximize CPU utilization and minimize resource waste.

Same as the previous experiment, the cluster used consists of 12 machines partitioned into two racks.  Since it does a round robin scheduling of executors among all workers/nodes, Storm's default scheduler will schedule executors on all the 12 machines regardless of the actual computation each executor will use. In contrast, by using R-Storm, the user can provide hints on how much CPU computation each instance of a component in a topology needs, and by doing so, R-Storm will not necessarily need to use all machines.  R-Storm will only use as many machines needed to satisfy user specified resource requirements.  For this experiment, we will supply R-Storm with the CPU usage requirements for each component/executor in the topology to demonstrate how R-Storm can achieve the same level of throughput while at the same time use only a portion of the total number of machines.

Figure~\ref{cpu_bound_results} displays the timeline graphs for the throughput of Linear, Diamond, and Star topologies. Figure~\ref{cpu_util_compare} displays a comparison of the average CPU utilization of machines used in the cluster when scheduling using Storm's default scheduler versus R-Storm. For the Linear topology, the throughput of a scheduling by R-Storm using 6 machines is similar to that of Storm's default scheduler using 12 machines.  
R-Storm creates schedulings that better utilizes the machines is the reason why R-Storm performs just as well as Storm's default scheduler even when R-Storm uses fewer machines  For the Linear topology, the average CPU utilization when using R-Storm is 69\% higher than when using Storm's default scheduler.  For the Diamond topology, a scheduling by R-Storm using 7 machines performs as well as a scheduling done by Storm's default scheduler using 12 machines.  For the Diamond topology, the average CPU utilization when using R-Storm is 91\% higher than when using Storm's default scheduler.  For the Star Topology, even when R-Storm was using half of the machines Storm's default scheduler used, R-Storm still had much higher throughput than Storm's default scheduler.  Since Storm's default scheduler does not take resource requirement and availability into account, a scheduling is created in which one of the machines in the cluster gets over utilized in computational resources and creates a bottleneck that  throttles the overall throughput of the Star topology.  The average CPU utilization of the cluster when using R-Storm is 350\% better than that of when using Storm's default scheduler.  When provided with accurate information about computational usage and availability, R-Storm is able to better utilize CPU resources (as seen Figure~\ref{cpu_util_compare}) in a cluster which results in being able to use fewer machines to produce to same level of performance as compared against Storm's default scheduler.  Another lesson learned from this experiment is that a topology's performance may not necessarily scale with the number of machines. Without adjusting the parallelism of components, a topology's throughput will reach a ceiling at which adding more machines will not improve performance. The performance of a topology is more closely related to what resources are need and scheduling a topology among unnecessary number of machines can also cause an increase in communication latency

\subsection{Yahoo Topologies: PageLoad and Processing Topology}

We obtained the layouts of two topologies in use at Yahoo! Inc. to evaluate the performance of R-Storm on actual topologies used in industry. The layout of the Page Load and Processing topologies are displayed in Figure~\ref{PageLoadTopologyLayout} and ~\ref{ProcessingTopologyLayout}.  These two topologies are used by Yahoo! for processing event-level data from their advertising platforms to allow for near real-time analytical reporting. Figure~\ref{YahooTopologiesResults} shows the experimental results for Page Load and the Processing topologies when using R-Storm and default Storm. As shown in the graphs, the scheduling derived using 
R-Storm performs considerably higher than a scheduling by default Storm. On average, the Page Load and Processing Topologies have 50\% and 47\% better overall throughput, respectively, when scheduled by R-Storm as compared to Storm's default scheduler.




\subsection{Multi-topology Performance}

In this section, we evaluate performance of R-Storm when scheduling multiple topologies in a cluster.  We compare the performance of R-Storm's schedulings to that of default Storm.  For this experiment, we used a larger 24 machine cluster separated into two 12 machine subclusters.  We submit both the Yahoo! PageLoad and Processing topologies to be scheduled by R-Storm and Default Storm.  Figure~\ref{muli-topology-throughput} displays the throughput comparison of R-Storm versus default Storm.  When scheduled by R-Storm, the throughput of both the PageLoad and Processing topologies (especially the Processing topology) are higher than when using default Storm.  When using default Storm, the performance of the Processing topology grinded to a near halt with an average overall throughput near zero.  The terrible performance of the Processing topology when using default Storm is another example of the consequences of over utilizing certain machines and not scheduling topologies in a resource-aware manner.

The average throughput of the PageLoad topology when scheduled by R-Storm (25496 tuples/10sec) is around 53\% higher than when scheduling by default Storm (16695 tuples/10sec).  For the Processing topology, the average throughput when scheduled using R-Storm (67115 tuples/10sec) is orders of magnitude higher than the throughput when scheduled using default Storm (10 tuples/sec).

\section{Related Work}
\label{relatedwork}  
Not much work has been done in the area of resource-aware scheduling in Storm or distributed data stream systems. Some research work has been done in the space of scheduling for Hadoop MapReduce which share many similarities with real-time processing distributed system. In the work from Jorda \textit{et al.}\cite{Polo:2011:RAS:2188461.2188475}, a resource-aware adaptive scheduling for a MapReduce Cluster is proposed and implemented. Their work is built upon the observation that there can be different workload jobs, and multiple users running at the same time on a cluster. By taking into account the memory and CPU capacities for each Task Tracker, the algorithm is able to find a job placement that maximizes a utility function while satisfying resource limits constraints. The algorithm is derived from a heuristic optimization algorithm to \textit{Class-Constrained Multiple Knapsack Problem} which is NP-hard. However, their work fails to take network into resource constraints which is a major bottleneck for streaming systems such as Storm.

Aniello \textit{et al.} in their paper\cite{Aniello} proposes two schedulers for Storm.  The first scheduler  is used in an offline manner prior to executing the topology and the second scheduler is used in on online fashion to potentially reschedule after a topology has been running for a duration.  Since R-Storm derives the scheduling prior to the actual execution of the topology, the offline scheduler proposed is the most comparable with the scheduling mechanism used in R-Storm. Their offline scheduler attempts to derive a linearization(similar to R-Storm) topology components and schedule tasks from those components in a round robin fashion to physical machines.  Their offline scheduler, like R-Storm, tries to minimize network distance between components that communicate with each other but uses a less effective approach and is limited to only acyclic Storm topologies which is not a limit for R-Storm.  Their online scheduler monitors CPU usage and aims to rebalance the topology to eliminate over utilization after running/profiling the topology for a duration of time.  However, their scheduler only takes into account the CPU usage of the nodes, with an average of 20-30 percentage improvement in performance in total. This work also does not take into account resource constraints and makes no attempt to satisfy or guarantee any resources that may be needed to run user defined code as does R-Storm. We attempted to deploy their code to profile its scheduling performance but the code was outdated and could run at all.
  
In the paper\cite{Wolf:2008:SOS:1496950.1496970}, Joel \textit{et al.} described a scheduler for System S, a distributed stream processing system which is similar to Storm, . The proposed algorithm is broken into four phases and runs periodically. At the first and second phases, the algorithm decides which job to admit, which job to reject, and compute the candidate processing nodes. In the third and last phases, it computes the fractional allocations of the processing elements to nodes. However, the approach only accounts processing power as resource and the algorithm itself is relatively complex requiring certain amount of computation.

\section{Conclusions}
\label{conclusion}
As an emerging open source technology in the field of Big Data, Storm has the capability of processing streams of data in a reliable and real time manner.  The growing popularity of Storm stems from the wide range of use cases for this platform.  Thus, we have created R-Storm to improve this emerging platform.
In spite of Storm's promising nature, the scheduling mechanism is inadequate. The schedulers that come with Apache Storm schedule tasks in a round-robin fashion with disregard to resource demands and availability. We designed and implemented a system called R-Storm that implements resource-aware scheduling within Storm. When scheduling tasks, R-Storm can satisfy both soft and hard resource constraints as well as minimizing network distance between components that communicate with each other.  We learn through R-Storm that round-robin scheduling with disregard to resource requirement and availability and communication patterns may be ineffective if not catastrophic.  Using user provided info and analyzing the topology DAG can lead to better initial schedulings.

We evaluate R-Storm by running a variety of micro - benchmark Storm topologies as well as production topologies used by Yahoo!. We compare the results of the topologies scheduled by R-Storm versus the default scheduler within Storm. Our experimental results demonstrate that schedulings done by R-Storm perform far better than that done by the default schedulers of Storm. From our experimental results we conclude that R-Storm achieves 30-47\% higher throughput and 69-350\% better CPU utilization than default Storm for the micro-benchmarks topologies. For the Yahoo! Storm topologies, R-Storm outperforms default Storm by around 50\% based on overall throughput. We also demonstrate that R-Storm performs much better when scheduling multiple topologies than default Storm.

The concepts and algorithms used in R-Storm are not only relevant for Storm but they are also applicable to other distributed data stream processing systems that has a DAG based data processing model (e.g. Twitter Heron \cite{kulkarni2015twitter}, Apache Flink \cite{flink_web}, etc) and should yield similar improvement as in Storm.

\bibliographystyle{abbrv}
\bibliography{stormRef}  
%
%


\balancecolumns
\end{document}